\documentclass[10pt,letterpaper]{article}
\usepackage{opex3}
\usepackage{graphicx}
\usepackage{subfigure}
\usepackage{epsfig}
\newcommand{\hypF}{\ensuremath{\,{}_1\!F_1}}
\begin{document}

\title{Improved focusing with Hypergeometric-Gaussian type-II optical modes}

\author{Ebrahim Karimi$^{1,2,*}$ , Bruno Piccirillo,$^{1}$ Lorenzo
Marrucci,$^{1,2}$ and Enrico Santamato$^{1,*}$}

\address{$^{1}$Dipartimento di Scienze Fisiche, Universit\`{a}
degli Studi di Napoli ``Federico II'', Complesso di Monte S. Angelo,
80126 via Cintia Napoli, Italy\\
$^{2}$Consiglio Nazionale delle Ricerche-INFM Coherentia,
Universit\`{a} di Napoli, Italy} \email{$^*$Corresponding authors:
karimi@na.infn.it, enrico.santamato@na.infn.it}
\begin{abstract}
We present a novel family of paraxial optical beams having a
confluent hypergeometric transverse profile, which we name
hypergeometric Gauss modes of type-II (HyGG-II). These modes are
eigenmodes of the photon orbital angular momentum and they have the
lowest beam divergence at waist of HyGG-II among all known finite
power families of paraxial modes. We propose to exploit this feature
of HyGG-II modes for generating, after suitable focusing, a ``light
needle'' having record properties in terms of size and aspect ratio,
possibly useful for near-field optics applications.
\end{abstract}

\section{Introduction}

In recent years, an increasing research effort has been put into
creating paraxial light beams having uncommon properties tailored
for particular uses. Notable examples include non-diffracting
beams~\cite{Dholakia08}, beams with large longitudinal non-
propagating component of the field~\cite{Wang08}, and beams
possessing an integer value of the photon orbital angular
momentum~\cite{molinaterriza07}. Such special beams have found a
wide range of applications, such as optical lithography, data
storage, microscopy, material processing, optical trapping, optical
tweezers and
metrology~\cite{Jeffries07,he95,durnin87,durnin87a,Zhao07,davis00,salo99}.

Richard and Wolf~\cite{Richard59} have shown, by using a vector
Debye integral, that a non-propagating component of the electric
field can be created near the focal point of high numerical aperture
lens. It has been found, both theoretically~\cite{Youngworth00} and
experimentally~\cite{Dorn03}, that a radially polarized light beam
can be focused into a much tighter and deeper spot than a linearly
polarized beam. One of the most interesting features of the radial
polarization is the formation of a large non-propagating
longitudinal component of optical electric field near the beam axis.
Conversely, the azimuthal polarization generates a strong magnetic
field near the optical axis. Recently, Zhan~\cite{Zhan06} has
studied the properties of circularly polarized vortex beams and has
found the proper combination of polarization and topological charge
of phase singularity to achieve focusing properties similar to
radial polarization. Besides polarization, other parameters such as
the pupil amplitude and phase structure of the field play an
important role to achieve a very narrow beam with a long depth at
focus, high beam quality and high optical efficiency. To this
purpose, many different optical beam profiles have been studied both
theoretically and experimentally, such as Bessel-Gauss (BG),
Hypergeometric (HyG), Hypergeometric-Gaussian (HyGG), fractional
elegant Laguerre-Gauss (fr-eLG), Ince-Gaussian, Laplace-Gauss and
Mathieu beams~\cite{bandres08,kotlyar07,karimi07}. Among them, the
radially polarized BG beams have been to date proved to provide the
best results. Moreover, Wang et al.~\cite{Wang08} have recently
calculated that an even tighter and deeper focus spot - a ``light
needle'' - even smaller than the standard diffraction limit, can be
obtained from BG beams by adding a suitable binary phase mask to the
high numerical aperture focusing system.

In this article, we present a new family of solutions to the
paraxial wave equation carrying finite power and having better
features than the BG beams under strong focusing. The modes studied
here have a hypergeometric amplitude profile as other modes
introduced previously (HyG~\cite{kotlyar07} and HyGG~\cite{karimi07}
modes), but differ from those in some important respects. For this
reason, we propose to name these new modes ``hypergeometric-gauss
modes of the second type'' (HyGG-II). Actually, these modes can be
also regarded as a limit subfamily of  circular beam (CiB)
introduced in \cite{bandres08}, where however, their special
features under focusing were not analyzed or discussed. We studied
the properties of the HyGG-II modes at the focus of a large aperture
lens. In this work, we prove that HyGG-II beams may provide better
spot size, beam quality and depth of focus than BG beams.

\section{Hypergeometric-Gaussian type-II modes}
The scalar Helmholtz paraxial wave equation in the cylindrical
coordinates is given by
\begin{eqnarray}\label{eq:pwe}
    \left(\partial_{\rho,\rho}+\frac{1}{\rho}\,\partial_\rho+\frac{1}{\rho^2}
    \,\partial_{\phi,\phi}+4\,i\partial_\zeta\right)\psi(\rho,\phi;\zeta)=0,
\end{eqnarray}
where $\rho=\frac{r}{w_{0}}$, $\phi$, $\zeta=\frac{z}{z_{R}}$ are
dimensionless cylindrical coordinates. Here $w_{0}$ is the beam
waist, $z_{R}=\frac{\pi\,w_{0}^2}{\lambda}$ is the beam Rayleigh
range and $\lambda$ is the beam wavelength. A family of solutions of
Eq.~(\ref{eq:pwe}) is given by
\begin{eqnarray}\label{eq:u}
    |\hbox{HyGG-II}\rangle_{pm}&\equiv& u_{pm}(\rho,\phi;\zeta)=
    C_{pm}\, \left(\frac{1}{1+i \zeta}\right)^{p/2+|m|+1}\nonumber\\
    &&\times\,\rho^{|m|}\,e^{-\frac{\rho^2}{(1+i\zeta)}}\,e^{i\,m\phi}\,
    \hypF\left(-\frac{p}{2},|m|+1;\frac{\rho^2}{(1+i\zeta)}\right),
\end{eqnarray}
where $m$ is an integer, $p$ is a real number, $\Gamma(x)$ is the
Gamma function, $\hypF\left(a,b;x\right)$ is the confluent
hypergeometric function and $C_{pm}$ is the normalization factor
given by $C_{pm}=\sqrt{\frac{2^{p+|m|+1}}{\pi\Gamma(p+|m|+1)}}\,
\frac{\Gamma\left(1+|m|+\frac{p}{2}\right)}{\Gamma\left(|m|+1\right)}$.
This solution can be obtained as a limit case of Eq. (6) of
\cite{bandres08} by setting $\gamma = -i(p+m+1)$, $q_0 =-i\,z_R$,
and $\tilde{q}_0\rightarrow\infty$. We see that $C_{pm}$ stays
finite as long as $p$ is so that $p\geq-|m|$, which ensures that the
power carried by HyGG-II beams is finite. Finally, the factor
$\exp(i m\phi)$ in Eq.~(\ref{eq:u}) ensures that HyGG-II beams are
eigenmodes of the photon orbital angular momentum. It can be shown
that the HyGG-II modes, like the HyGG modes~\cite{karimi07} (but
unlike the HyG modes~\cite{kotlyar07}), form a non-orthogonal set,
i.e. ${}_{p'm'}\langle
\hbox{HyGG-II}|\hbox{HyGG-II}\rangle_{pm}=\delta_{mm'}
\frac{\Gamma(p/2+p'/2+|m|+1)}{\sqrt{[\Gamma(p'+|m|+1)\Gamma(p+|m|+1)]}}$.
Moreover, the HyGG-II modes, unlike the HyGG and HyG ones, have no
singularity at $\zeta=0$.
\begin{figure}[!htbp]
    \begin{center}
    \includegraphics[width=.8\textwidth]{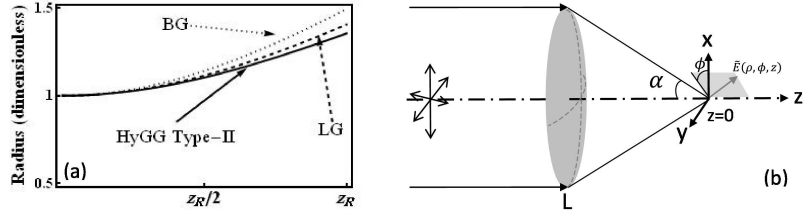}
     \end{center}
     \caption{(a) Variation of maximum intensity radius during the propagation
    for LG$_{0,1}$, BG$_{1}$, HyGG-II$_{-1,1}$. (b) Scheme of the focusing system.}\label{fig:1}
\end{figure}
A very interesting property of HyGG-II modes is that they suffer
very low diffraction. In fact, the divergence angle at waist of the
HyGG-II modes is smaller than of the BG and LG modes.
Fig.~\ref{fig:1}(a) compares the variation of the maximum intensity
radius of HyGG-II, BG and LG beams while they propagate along the
$z$-axis. We see that HyGG-II has a slope 12.5\% lower than LG beam
and 28\% lower than BG beam, where we define
$\hbox{slope}=\frac{w(z_{R})-w_{0}}{z_{R}}$. The HyGG-II modes may
be generated from a plane-wave by the following unbounded
transmittance function at $\zeta=0$
\begin{eqnarray}\label{eq:pupil}
    u_{pm}(\rho,\phi,0)=e^{-\rho^2+i\,m\phi}\,\rho^{|m|}\,\hypF\left(-\frac{p}{2},|m|+1;\rho^2\right).
\end{eqnarray}
The asymptotic behavior of HyGG-II at large $\rho$ is given by
\begin{eqnarray}\label{eq:asym}
    u_{pm}(\rho,\phi,0)\propto e^{-\frac{\rho^2}{1+i\zeta}}\,\rho^{p+|m|}+
    \frac{\rho^{-(p+|m|+2)}}{\Gamma(-\frac{p}{2})}.
\end{eqnarray}
Therefore, there are two asymptotic behaviors for the intensity
$I_{p,m}=|u_{p,m}|^2$ depending on the radial number $p$;
\begin{enumerate}
  \item In the general case, $I_{p,m}\propto\rho^{-2(p+|m|+2)}$, i.\ e.\ a power law decay.
  \item For nonnegative even integer number $p$, $I_{p,m}\,\propto\rho^{2(p+|m|)}
  \,e^{-\frac{2\rho^2}{1+\zeta^2}}$, i.\ e.\ a gaussian decay.
\end{enumerate}
Because all zeros of the hypergeometric function are on the real
axis, the intensity of the HyGG-II modes never vanishes in the
transverse plane for $\zeta>0$, except at $\rho=0$ where the vortex
singularity is located when $m\neq0$.
Finally, we will discuss briefly some possible subfamilies of the
HyGG-II modes.
\begin{itemize}
  \item [a)] $p=|m|=0$, in this case the mode is the gaussian TEM$_{00}$ beam.
  \item [b)] $p=-|m|$ a negative integer number, the HyGG-II can be expanded as
  a superposition of two modified-Bessel beams;
\begin{eqnarray}\label{eq:sub1}
    &&u_{-|m|,m}(\rho,\phi;\zeta)=\frac{1}{\sqrt{2}}\,\left(\frac{1}{1+i\zeta}\right)^{3/2}\,e^{i\,m\phi}\,
    \rho\,e^{-\frac{\rho^2}{2(1+i\zeta)}}\cr && {\times}\,\left[I_{\frac{|m|-1}{2}}\left(\frac{\rho^2}{2(1+i\zeta)}\right)-
    I_{\frac{|m|+1}{2}}\left(\frac{\rho^2}{2(1+i\zeta)}\right)\right]
\end{eqnarray}
where $I_{n}(x)$ is the modified Bessel function of integer order
$n$. We call this sub-family ``modified Bessel Gauss modes of
type-II'' (MBG-II), for distinguishing them from those introduced
in~\cite{karimi07}.
  \item [c)] For $p\geq 0$ even integer number, one can easily show that
\begin{eqnarray}\label{eq:sub2}
    u_{2n,m}(\rho,\phi;\zeta)&=& \sqrt{\frac{2^{2n+|m|+1}}{\pi \Gamma(2n+|m|+1)}}\,\Gamma(n+1)
    \,\left(\frac{1}{1+i\zeta}\right)^{n+|m|+1}\nonumber\\
    &&{\ensuremath\times}e^{-\frac{\rho^2}{1+i\zeta}}\,\rho^{|m|}\,e^{i\,m\phi}\,L_{n}^{|m|}(\frac{\rho^2}{1+i\zeta}).
\end{eqnarray}
which are the well-known ``elegant Laguerre-Gauss'' (eLG)
beams~\cite{bandres08}.
  \item [d)] For $p>0$ odd integer number, the HyGG-II modes reduce
 to a polynomial superposition of the modified Bessel functions $I_0(x)$ and $I_1(x)$.
 We will call this sub-family ``modified-polynomial
 Bessel-Gauss'' (MPBG) beams.
\end{itemize}
\section{Hypergeometric-Gaussian-II modes under strong focusing}
In this section, we study some properties of HyGG-II in the focal
region of a high numerical aperture lens. Let us consider an
aplanatic high-numerical-aperture focusing lens system (see
Fig.~\ref{fig:1}(b). The origin $z=0$ is located at the lens focal
point, $f$, NA, $\alpha=\arcsin{(\frac{\hbox{NA}}{n})}$, and $n=1$
are the focal length, the numerical aperture, the semi-aperture
angle, and the vacuum refractive index, respectively. Using the
vectorial Debye diffraction integral, Richard and
Wolf~\cite{Richard59} have shown that the electric field at the
point $\widetilde{r}=(\rho,\phi,z)$, in a region close to the focal
point in the cylindrical coordinates is given by
\begin{eqnarray}\label{eq:debye}
    \widetilde{E}(\widetilde{r})=-\frac{i}{\lambda}\int\!\!\int_{\Omega}
    \widetilde{a}(\theta,\varphi)e^{2\pi i\left(z\cos{\theta}+\rho\sin{\theta}
    \cos{(\varphi-\phi)}\right)}d\Omega,
\end{eqnarray}
where $\widetilde{a}(\theta,\varphi)$ is determined by the field
distribution in the object space at the pupil and $\Omega$ is the
solid angle. In Eq.~(\ref{eq:debye}), $\rho$ and $z$ are
dimensionless, their scale length being the $\lambda$. Youngworth
and Brown~\cite{Youngworth00} have calculated this integral for
radially and azimuthally polarized beams. Their calculation showed
that the radial polarization is much more effective than the
azimuthal and linear ones for obtaining a tight focusing. Therefore,
we restrict our attention only to radially polarized beams. The
electric field of a radially polarized beam in the focal region is
given by
\begin{eqnarray}\label{eq:radially}
    E_{\rho}(\widetilde{r})=\frac{f}{\lambda}\int_{0}^{\alpha}\sqrt{\cos{\theta}}
    \sin{(2\theta)}\,l(\theta)\,J_{1}(2\pi\rho\sin{\theta})e^{i(2\pi z\cos{\theta})}d\theta\cr
    E_{z}(\widetilde{r})=\frac{2if}{\lambda}\int_{0}^{\alpha}\sqrt{\cos{\theta}}
    \sin^{2}{\theta}\,l(\theta)\,J_{0}(2\pi\rho\sin{\theta})e^{i(2\pi
    z\cos{\theta})}d\theta,
\end{eqnarray}
where $J_{0}(x)$ and $J_{1}(x)$ are Bessel's functions, and
$l(\theta)$ is the amplitude distribution of the pupil apodization
function.
\begin{figure}[!htbp]
    \begin{center}
    \includegraphics[width=.8\textwidth]
    {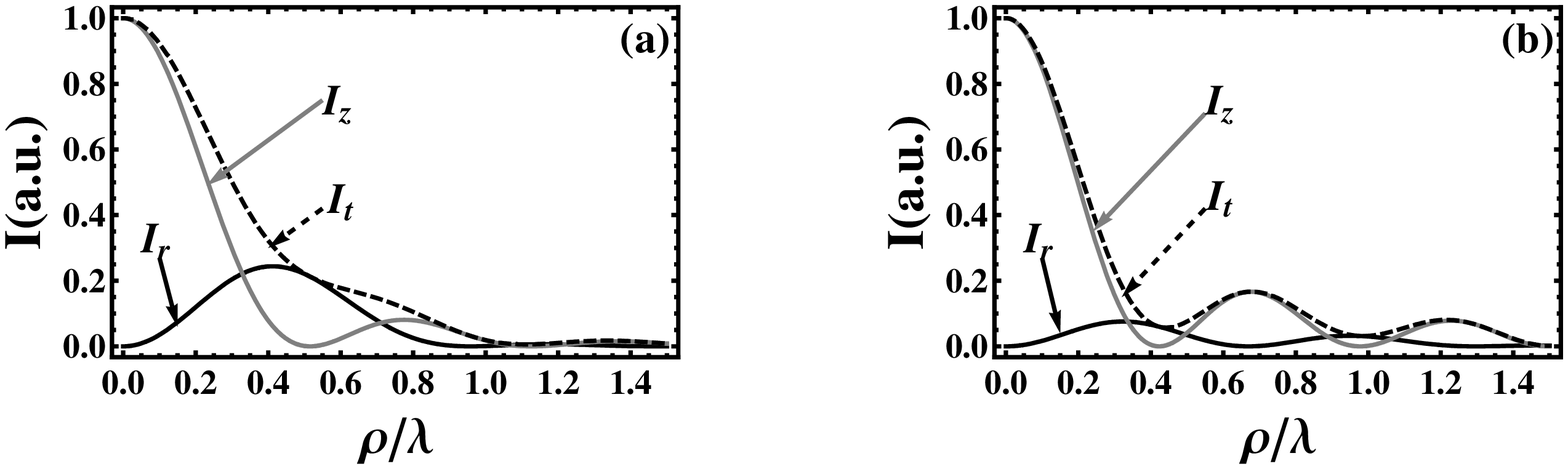}
    \end{center}
  \caption{Intensity profile of the longitudinal and radial field components at the
  focal point of a high numerical aperture lens (a) and a system of high numerical
  aperture lens and binary phase mask (b). The black, gray  and dashed lines are radial,
   longitudinal and total intensity, respectively.}\label{fig:2}
\end{figure}
We consider an amplitude-only apodization function given by the
HyGG-II$_{-1,1}$ profile calculated from Eq.~(\ref{eq:pupil}) with
the singular phase factor omitted. We choose this profile because it
exhibits minimum diffraction. Explicitly, $l(\theta)$ is given by
\begin{equation}\label{eq:apodiz}
    l(\theta)=e^{-\beta^2\left(\frac{\sin{\theta}}{\sin{\alpha}}\right)^2}\,
    \left(\beta\frac{\sin{\theta}}{\sin{\alpha}}\right)
    \hypF\left(\frac{1}{2},2;\beta^2\left(\frac{\sin{\theta}}{\sin{\alpha}}\right)^2\right),
\end{equation}
where $\beta$ is the ratio between the pupil radius and the beam
waist. One can solve the integral in Eq.~(\ref{eq:apodiz})
numerically and plot the intensities near the focus for any values
of the numerical aperture. The numerical integrations were performed
using global adaptive algorithms, which recursively subdivide the
integration region as needed. In our calculations we considered
$\beta=1$ and $\hbox{NA}=0.95$ corresponding to
$\alpha=71.8^{\circ}$. We compared our results with the well-known
apodization function of the BG$_1$ beam. Fig.~\ref{fig:2}(a) shows
the intensity profile of the radial and longitudinal components of
the optical field at focus. It is evident that the intensity of the
longitudinal component is higher than the radial component. The beam
quality is characterized by~\cite{Wang08}
\begin{equation}\label{eq:eta}
    \eta=\frac{\int_{0}^{r_{0}}|E_{z}(r,0)|^2 r\,dr}{\int_{0}^{r_{0}}|E_{\rho}(r,0)|^2 r\,dr+
    \int_{0}^{r_{0}}|E_{z}(r,0)|^2 r\,dr}
\end{equation}
where $r_0$ is the first zero of the radial field component. For the
HyGG-II$_{-1,1}$ profile we found $\eta=52.5$\% instead of 44.7\%
for the BG$_1$ profile in the same conditions. The hypergeometric
apodization function increases the beam quality by 17\%.
Furthermore, the HyGG-II$_{-1,1}$ beam size (full width at half
maximum, FWHM) is as small as $0.60\lambda$ which is 13\% less than
the beam size of the BG$_1$ mode (although it is still larger than
diffraction limit). Also, its depth focus is about $\sim1.5\lambda$
which is 1.4\% larger than the depth focus of the BG$_{1}$ mode
\begin{figure}[!htbp]
    \begin{center}
    \includegraphics[width=0.56\textwidth]
    {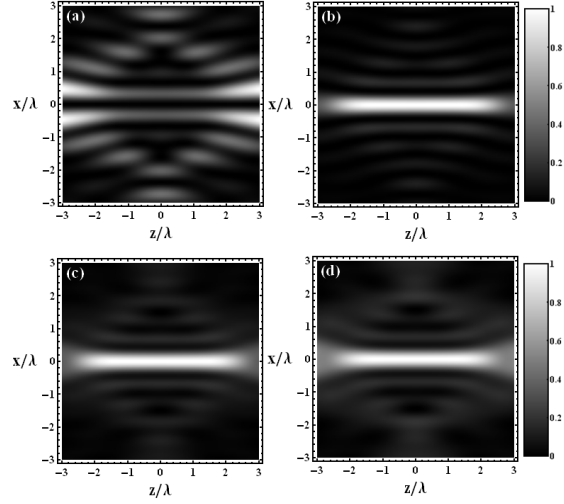}
    \end{center}
  \caption{Density plots of intensity distribution for (a) radial, (b) longitudinal components. (c) and (d)
  show the total intensity distribution for the HyGG-II$_{-1,1}$ and the BG$_1$ beams, respectively.}\label{fig:3}
\end{figure}
Finally, we calculated the optical electric field near the focal
point when a binary phase mask is inserted just in front of the high
numerical aperture system, as in~\cite{Wang08}. Our binary phase
mask is made of five concentric belts in which the phase of each
belt changes by $\pi$ with respect to neighbor
belts~\cite{Wang06,Wang08}. Compared with the mask used
in~\cite{Wang08} for the BG$_1$ modes, we introduced some small
modifications, i.e.,
$\theta_{1}=4.46^{\circ},\,\,\theta_{2}=23.64^{\circ},\,\,\theta_{3}
    =36.53^{\circ},\,\,\theta_{4}=49.03^{\circ}$,
where $\theta_{i}$ are related to the inner radius of each belts,
$r_{i}=\sin{\theta_{i}}/\hbox{NA}$. The intensity profiles of the
radial and longitudinal components of the field in the presence of
this phase mask are shown in Fig.~\ref{fig:2}(b). The total field
FWHM is $0.426\lambda$, leading to a spot size of $0.142\lambda^2$.
These values are 1.6\% and 3.2\% smaller than the in BG$_1$
case~\cite{Wang08}. Furthermore, the phase mask increases the beam
quality to 81.76\%, which is improved by 1.7\% with respect to the
BG$_1$ profile. Fig.~\ref{fig:3} shows the intensity distribution
for the radial Fig.~\ref{fig:3}(a), longitudinal Fig.~\ref{fig:3}(b)
and total field Fig.~\ref{fig:3}(c) of the HyGG-II$_{-1,1}$ profile.
For the sake of comparison, the total field of the BG$_1$ profile is
shown in Fig.~\ref{fig:3}(d). It is clear that the depth of focus of
the hypergeometric profile is longer than the Bessel-Gauss case and
it is about $4.5\lambda$.

It is also worth investigating the field distribution obtained at
the focus when using vector vortex beams~\cite{Zhan06} having the
HyGG-II profile of Eq.~(\ref{eq:pupil}) with the singular phase
retained, but we postponed this problem to the near future.
\section{Conclusion}
In conclusion, we studied a novel family of paraxial beams having
singular profile. All modes of this family are eigenmodes of the
orbital angular momentum of light and are an overcomplete
nonorthogonal set of modes. Moreover, the modes of this family
exhibit a diffraction divergence at waist which is the smallest
among all other known modes carrying a finite power. As a
consequence, we found that these beams, when radially polarized, can
be focused into a tighter and longer region, compared with other
known modes. A small improvement compared with other modes is
obtained also when the focusing is pushed below the diffraction
limit by adding a suitable phase mask.
\end{document}